\documentstyle[preprint,eqsecnum,aps,epsf]{revtex}

\begin{document}
\draft
\preprint{HEP/123-qed}
\title{Heat-capacity anomalies at $T_{sc}$ and $T^{*}$ in the ferromagnetic superconductor UGe$_2$}

\author{N. Tateiwa,$^{1,2,}$\cite{byline1} T. C. Kobayashi,$^{1,}$\cite{byline2}  K. Amaya,$^{2}$ Y. Haga,$^{3}$ R. Settai,$^{4}$ and Y. \=Onuki$^{3,4}$}
\address{
$^{1}$Research Center for Materials Science at Extreme Conditions, Osaka University, Toyonaka, Osaka 560-8531, Japan\\
$^{2}$Department of Physical Science, Graduate School of Engineering Science, Osaka University, Toyonaka, Osaka 560-8531, Japan\\
$^{3}$Advanced Science Research Center, Japan Atomic Energy Research Institute, Tokai, Ibaraki 319-1195, Japan\\
$^{4}$Department of Physics, Graduate School of Science, Osaka University, Toyonaka, Osaka 560-0043, Japan
}
%\date{\today}
\date{30 December 2003}
\maketitle
\begin{abstract}
 The heat-capacity and magnetization measurements under high pressure have been carried out in a ferromagnetic superconductor UGe$_2$. Both measurements were done using a same pressure cell in order to obtain both data for one pressure. Contrary to the heat capacity at ambient pressure, an anomaly is found in the heat capacity at the characteristic temperature $T^{*}$ where the magnetization shows an anomalous enhancement under high pressure where the superconductivity appears. This suggests that a thermodynamic phase transition takes place at $T^{*}$ at least under high pressure slightly below $P_{c}^{*}$ where $T^{*}$ becomes zero. The heat-capacity anomaly associated with the superconducting transition is also investigated, where a clear peak of $C/T$ is observed in a narrow pressure region ($\Delta P \sim 0.1$ GPa) around $P_{c}^{*}$ contrary to the previous results of the resistivity measurement. Present results suggest the importance of the thermodynamic critical point $P_{c}^{*}$ for the appearance of the superconductivity. 
\end{abstract}

 \narrowtext
  Recently the pressure-induced superconductivity was found in UGe$_2$.~\cite{rf:saxena,rf:huxley1} This finding is quite interesting since the superconductivity appears in the pressure range from 1.0 to 1.6 GPa where UGe$_2$ is still in the ferromagnetic state. This is the first case where the same 5{\it f} electrons are involved with both orderings.

 From the theoretical point of view, the coexistence of superconductivity and ferromagnetism has been considered over several decades.~\cite{rf:fay,rf:kirkpatrick,rf:roussov} Basically these theories  assume that the superconductivity is mediated by the low-energy magnetic excitation which is enhanced near the ferromagnetic quantum critical point (QCP) where a second-order phase transition is driven to 0 K by the parameter such as the pressure or the stoichiometric composition of a sample. The superconductivity is predicted to appear in both ferromagnetic and paramagnetic sides of the QCP. However, contrary to the above theoretical predictions, the mechanism of superconductivity in UGe$_2$ seems to have no relation with the ferromagnetic fluctuation associated with the ferromagnetic critical point $P_{c}$ since the transition from the ferromagnetic state to the paramagnetic state is established as a first order in UGe$_2$.~\cite{rf:huxley2,rf:settai,rf:terashima1}

 In UGe$_2$, there is another "phase boundary" at the characteristic temperature $T^{*}$ in the ferromagnetic ordered state below $T_{\rm C}$. At $T^{*}$, there is a broad anomaly in the resistivity and the thermal expansion.~\cite{rf:oomi} The magnetization shows an anomalous increase below $T^{*}$.~\cite{rf:huxley1,rf:tateiwa1,rf:pfleiderer,rf:motoyama} Experimentally the microscopic origin of $T^{*}$ is not clear at present. With increasing pressure, $T^{*}$ decreases monotonously and becomes 0 K at a pressure $P_{c}^{*}$ ($\sim$ 1.20 GPa). The superconducting transition temperature $T_{sc}$ shows a maximum value around $P_{c}^{*}$.~\cite{rf:huxley1,rf:tateiwa2} Thus it was pointed out that the superconductivity was mediated by the low-energy magnetic excitation around the "critical point" $P_{c}^{*}$.~\cite{rf:saxena,rf:huxley1} This point of view implicitly assumes the $P_{c}^{*}$ as a second-order QCP. However there is no distinct anomaly around $T^{*}$ in the heat capacity at ambient pressure.~\cite{rf:huxley1,rf:onuki} Thus, for the understanding of the coexistence of the superconductivity and the ferromagnetism in UGe$_2$, it is highly desirable to clarify that a real transition occurs at $T^{*}$ and that $P_{c}^{*}$ corresponds to a thermodynamic critical point from the heat-capacity measurement.

  In this Rapid Communication, we report the result of the heat-capacity measurement on UGe$_2$ under high pressure. Our experimental results suggest that a second-order phase transition takes place at $T^{*}$ at least in a high-pressure region. The heat-capacity anomaly associated with the superconducting transition is observed in the narrow pressure region around $P_{c}^{*}$.

   A single crystal was grown by the Czochralski pulling method in a tetra-arc furnace as described in Ref. 13. The residual resistivity ratio was 600 at ambient pressure, indicating high quality. The Cu-Be piston-cylinder cell was designed as can be used in the heat-capacity and magnetization measurements so that both data were obtained for the same pressure. The heat capacity $C(T)$ was measured by the adiabatic heat pulse method using a $^{3}$He-$^{4}$He dilution refrigerator, while the magnetization measurement was done by using a commercial superconducting quantum interference device magnetometer.

 Figure 1 (a) shows the temperature dependence of $C/T$ at 1.15 ($< P_{c}^{*}$) and 1.28 GPa ($> P_{c}^{*}$) under zero magnetic field. The temperature dependence of the magnetization under the magnetic field of 0.5 T is also plotted. At 1.15 GPa, the magnetization shows a characteristic increase below $T^{*}\sim6$ K. Correspondingly a heat-capacity anomaly is found to appear around $T^{*}$, which is contrary to the heat capacity at ambient pressure where a distinct anomaly is absent around $T^{*}$. This observation suggests that a phase transition takes place at $T^{*}$ at least under high pressure just below $P_{c}^{*}$. The transition temperature is defined as $T^{*}$ = 6.0 K such that the entropy is conserved as drawn by a broken line in Fig. 1 (a). The value of $\Delta C/T^{*}$ is about 70 $\pm$ 20 mJ/mol K$^2$, where $\Delta C$ is the jump of the heat capacity at $T^{*}$. At 1.28 GPa, above $P_{c}^{*}$, there is no anomaly in both the heat capacity and the magnetization. The temperature dependence of the entropy $S(T)$ obtained by the $C/T$ curve at 1.15 GPa ($< P_{c}^{*}$) deviates from that of 1.28 GPa ($> P_{c}^{*}$) below $T^{*}$ due to the phase transition at $T^{*}$, while both $S(T)$ curves show good accordance above $T^{*}$. The magnetization at 1.15 GPa increases gradually with decreasing temperature below $T^{*}$. The curve of the heat capacity is of the $\lambda$-type. These results suggest the second-order phase transition at $T^{*}$. Therefore, the enhancement of the linear heat-capacity coefficient $\gamma_{n}$ seems to originate from the low-energy fluctuation around $P_{c}^{*}$ which causes the superconductivity. 

 The value of $\Delta C/T^{*}$ is roughly consistent with the recent thermal expansion ($\alpha_V$) result ($\Delta C/T^{*} \sim$ 55 mJ/mol K$^2$ ) estimated by the relation $dT^{*}/dP = ({\Delta\alpha}_V/(\Delta C/T^{*})$ at 1.02 GPa, assuming a second-order transition. Here, $dT^{*}/dP$ is the pressure dependence of $T^{*}$. $\Delta\alpha_V$ is the jump of $\alpha_V$ at $T^{*}$.~\cite{rf:ushida}

 It is revealed by the magnetization measurement that $T^{*}$ is induced by the application of the external field along an easy {\it a} axis in the pressure region above $P_{c}^{*}$.~\cite{rf:huxley1,rf:tateiwa1,rf:pfleiderer} In the present experiment, the field-induced anomaly of the heat capacity is observed at 1.28 GPa as shown in Fig. 1(b) where the magnetization starts to increase at $T^{*}$. 
Therefore, it is suggested that the inducement of $T^{*}$ is a thermodynamic phase transition. From these results, the magnetic-field dependence of $\gamma_{n}$ and $T^{*}$ is obtained as shown in Fig. 2 where the magnetization process at 1.8 K is also plotted.  The transition temperature $T^{*}$ is determined by the heat-capacity and the magnetization measurements as denoted by dotted line in Fig. 1. Around $H^{*}\sim$ 1.8 T where $T^{*}$ appears, the metamagnetic-like transition occurs in the magnetization process and correspondingly the $\gamma_{n}$ value decreases drastically. The $\gamma_{n}$ value tends to saturate at higher magnetic fields. This suggests that the mass of the quasiparticles is strongly enhanced around the phase boundary where $T^{*}$ becomes 0 K. The magnetic-field dependence of $\gamma_{n}$ is qualitatively consistent with the recent results of de Haas-van Alphen (dHvA) experiments above $P_{c}^{*}$ where the Fourier spectra of the dHvA oscillations taken below and above $H^{*}$ are different from each other and the cyclotron masses of all the branches detected above $H^{*}$ are lighter than those below $H^{*}$.~\cite{rf:haga,rf:terashima2} The enhancement of the magnetization below $T^{*}$ or the increment of $T^{*}$ by applying magnetic field can be phenomenologically understood from the thermodynamic point of view based on Ehrenfest's theorem which should be satisfied at the second-order phase transition:$$\Delta (\it\partial{M}/\partial{T})_{H} = -({\rm\Delta} \it C/T^{*})[dT^{*}/d(\mu_{\rm 0}H)].\eqno{(1)}$$ Since $dT^{*}/dH$ is positive in the measured field,~\cite{rf:tateiwa1} $\Delta \it(\partial{M}/\partial{T})_{H}$ should be negative, which is qualitatively consistent with the experimental result. The values of $\Delta C/T^{*}$ are about 70 $\pm$ 10 and 77 $\pm$ 10 mJ/mol K$^2$ at 2.5 and 3.0 T, and then the values of $\Delta \it(\partial{M}/\partial{T})_{H}$ at 2.5 and 3.0 T are estimated as -0.056 $\pm$ 0.007 and -0.044 $\pm$ 0.006 $\mu_{\rm B}/{\rm UK}$, respectively using Eq. (1). These values are in good agreement with those of -0.052 $\pm$ 0.007 and -0.039 $\pm$ 0.007 $\mu_{\rm B}/{\rm UK}$ at 2.5 and 3.0 T, respectively, shown by the dotted line in Fig. 1(b). This agreement suggests the second order phase transition at $T^{*}$. Watanabe and Miyake develop a microscopic theory assuming that $T^{*}$ is a coupled charge- and spin-density (CDW/SDW) transition temperature.~\cite{rf:watanabe} Various anomalous experimental results are explained by the theory. However, there is no experimental evidence for the CDW state at present. Recently, Pfleiderer and Huxley suggest that $P_{c}^{*}$ is a first-order critical point from the pressure dependence of magnetization at 2.0 K.~\cite{rf:pfleiderer} From the present study, it is clear that the enhancement of $\gamma$ is due to the critical fluctuation related to the phase transition at $T^{*}$. Therefore $P_{c}^{*}$ is considered to be the weakly first-order or the second-order critical points. Further experimental investigations are needed in order to understand the microscopic origin of the phase transition at $T^{*}$. 

  The temperature dependence of the heat capacity at low temperatures is shown in the form of $C/T$ in Fig. 3. At 1.22 GPa, a clear peak associated with the superconducting transition was observed at around 0.6 K, while the anomaly smears at 1.15 and 1.28 GPa. At 1.22 GPa, the transition temperature is estimated as $T_{sc}$ = 0.60 $\pm$ 0.10 K such that the entropy is conserved as drawn by a broken line in Fig. 3. The value of ${\Delta} C/(\gamma_{n} T_{sc})$ is 0.29 $\pm$ 0.06 where ${\Delta} C$ is the jump of the heat-capacity at $T_{sc}$ and $\gamma_{n}$ is the value of $C/T$ just above $T_{sc}$. The residual $\gamma$ value obtained by the extrapolation of $C/T$ curve linearly to 0 K as shown in Fig. 3 is $\gamma_{0}\,\sim$ 72 $\pm$ 5 mJ/mol K$^2$ which is about 70\% of $\gamma_{n}$. By a similar estimation for other data on this sample (no. 1), the pressure dependence of ${\Delta} C/(\gamma_{n} T_{sc})$ and $\gamma_{n}$ is obtained as shown by circles in Fig. 4. The experimental result on another sample (no. 2) which was cut from the same ingot for sample no. 1 is also shown by squares. The pressure dependence of $T_{sc}$ determined by zero resistance in the resistivity measurement using the sample with similar quality to the present sample is also plotted. The resistivity shows the superconductivity in a wide pressure range from 1.0 GPa to $P_{c}$ ($\sim$ 1.5 GPa). On the other hand, $\Delta C/(\gamma_{n} T_{sc})$  shows a maximum around $P_{c}^{*}$, and it is strongly suppressed when the pressure deviates from $P_{c}^{*}$. The maximum of $T_{sc}$ ($\sim$ 0.72 K) at around $P_{c}^{*}$ in the resistivity measurement is consistent with the temperature where $C/T$ starts to increase.
  
  The heat capacity reflects the bulk nature of a sample, while the resistivity is governed by the supercurrent through the paths of pure parts in a sample. Assuming that the zero resistivity indicates an ideal superconducting transition temperature $T_{sc0}$ expected for a sample without an impurity, the reduction rate of $T_{sc}$ ( = $T_{sc}$/$T_{sc0}$) due to the impurity is about 0.83 around $P_{c}^{*}$. The large $\gamma_{0}$ value indicates a large residual density of states at the Fermi energy $E_{\rm F}$. The phenomenological theories suggest that only the Fermi surface of the majority band opens the superconducting gap and the minority band remains a normal state below $T_{sc}$.~\cite{rf:machida,rf:fomin} Recent band calculations pointed out that the contribution to the density of states at $E_{\rm F}$ from the minority band is less than 10\% of the total density of states.~\cite{rf:shick,rf:yamagami} Thus it is not appropriate to ascribe the contribution from the minority band to the origin of the large $\gamma_{0}$ value. The contribution from the considerable self-vortex state due to the coexistence of the ferromagnetism and the superconductivity is also negligibly small because the distance of the inter-vortices ($\sim$ 1100 {\AA}), estimated from the spontaneous magnetization (0.19 T : $\mu_{\it ord}\sim1.0 \,\mu_{\rm B}/{\rm U}$ at $P_{c}^{*}$) assuming that vortices form the Abrikosov triangle lattice, is about ten times larger than the size of a vortex ($\sim$ the coherence length $\xi$, 130 {\AA}).~\cite{rf:tateiwa2,rf:tachiki,rf:abrikosov} It is well known that the small amount of impurity easily gives rise to a finite residual density of states at $E_{\rm F}$ in the superconductor with an anisotropic gap.~\cite{rf:schmitt} In the case of a triplet superconductor Sr$_2$RuO$_4$,~\cite{rf:maeno} the values of $\gamma_{0}$ and $T_{sc}$ are known to be very sensitive to a small amount of impurity, where the relation between the $\gamma_{0}$/$\gamma_{n}$ value and the reduction rate of $T_{sc}$ (= $T_{sc}$/$T_{sc0}$) were explained by the theory which evaluated the impurity effect on the $p$-wave superconductor treating the impurity scattering close to the unitarity limit.~\cite{rf:mao,rf:nishizaki} In the case of the present sample, the mean free path $l$ is about 1400 {\AA} determined by the dHvA experiment and then $l/\xi\,\sim$ 11.~\cite{rf:haga} The $\gamma_{0}$/$\gamma_{n}$ value and the reduction rate of $T_{sc}$ are estimated as $\gamma_{0}$/$\gamma_{n}$ $\sim$ 0.7 and $T_{sc}/T_{sc0}$ = 0.83. The application to the theory in Sr$_2$RuO$_4$ reveals $\gamma_{0}$/$\gamma_{n}$ = 0.45 $\pm$ 0.10 and $T_{sc}/T_{sc0}$  = 0.85 $\pm$ 0.05 for $l/\xi$ = 11.~\cite{rf:nishizaki,rf:kikugawa} The values of $T_{sc}/T_{sc0}$ for both compounds are in agreement within an experimental error. The larger value of $\gamma_{0}$/$\gamma_{n}$ in the case of UGe$_2$ might indicate that the present superconductivity is extremely sensitive to a small amount of impurity. These results suggest the existence of an anisotropic gap in superconductivity. 

  In summary, contrary to the absence of the anomaly in the heat capacity at ambient pressure, the heat-capacity anomaly is found at the characteristic temperature $T^{*}$ = 6.0 K at 1.15 GPa ($< P_{c}^{*}$). The thermodynamic consideration suggests that a second-order phase transition takes place at $T^{*}$ and $P_{c}^{*}$ is the weakly first-order or the second-order critical points. We also investigated the superconducting heat-capacity anomaly. The pressure dependence of ${\Delta} C/(\gamma_{n} T_{sc})$ suggests that the bulk superconducting phase exists in the narrow pressure region around $P_{c}^{*}$.~\cite{rf:sandeman}

 We thank S. Watanabe, K. Miyake, H. Yamagami, T. Terashima, K. Machida, H. Kotegawa, Y. Kitaoka, S. S. Saxena, M. J. Steiner, A. Huxley, I. Sheikin, J. Flouquet, G. Oomi, G. Motoyama, T. Nishioka, and N. K. Sato for fruitful discussions. One of authors (N.T.) thanks N. Kikugawa and Y. Maeno for sending their results on Sr$_2$RuO$_4$. This work was financially supported by the Grant-in Aid for COE Research (Grant No. 10CE2004) of the Ministry of Education, Science, Sports and Culture, and by CREST, Japan Science and Technology Corporation.

\begin{figure}
\begin{center}
  \epsfxsize=11cm
  \epsfbox{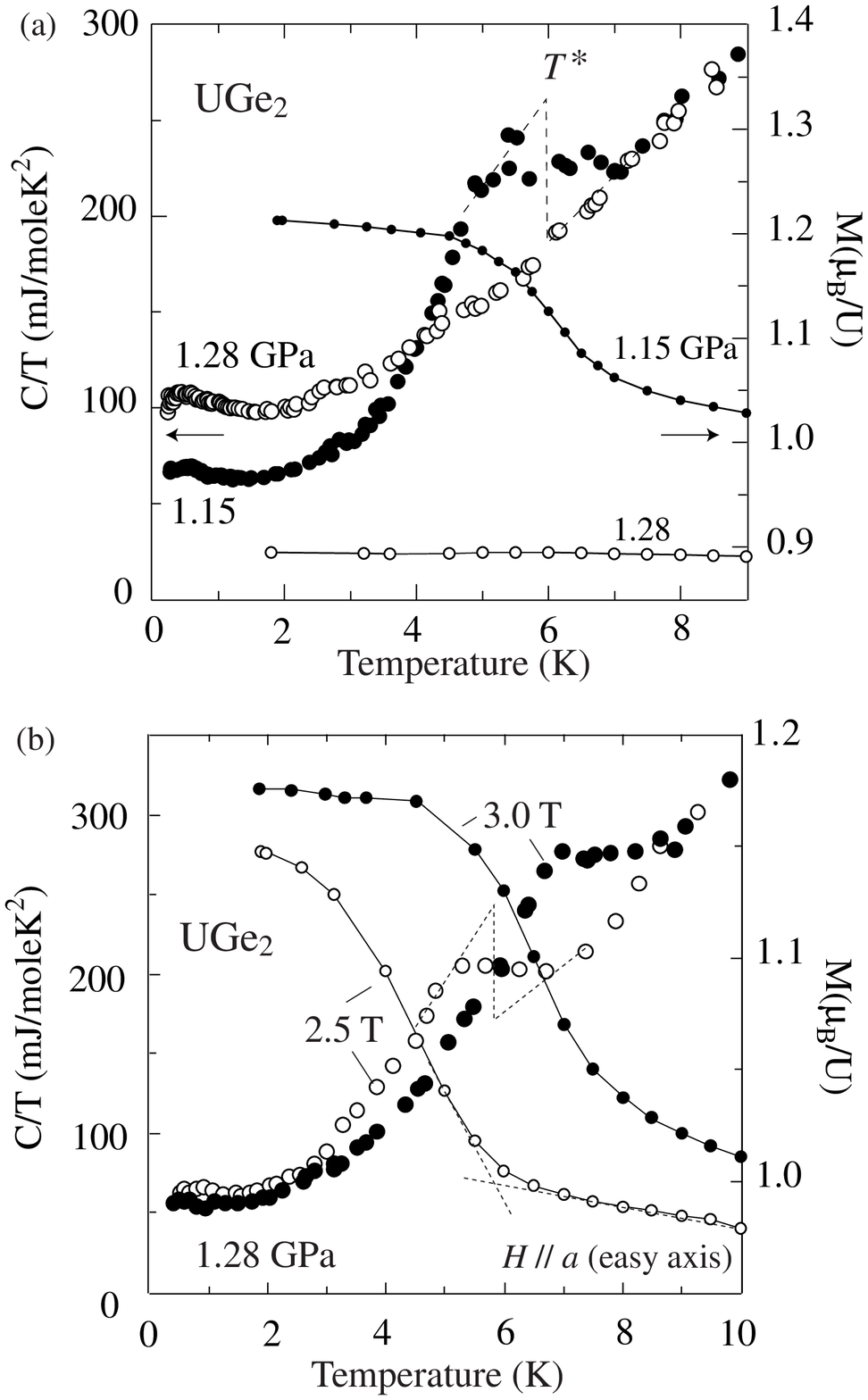}
\end{center}
\caption{ (a) Temperature dependence of $C/T$ for 1.15 and 1.28 GPa under zero magnetic field. The data connected by a line indicate the temperature dependence of the magnetization under 0.5 T. (b) Temperature dependence of $C/T$ under magnetic field of 2.5 and 3.0 T for 1.28 GPa. The data connected by a line indicate the temperature dependence of the magnetization under 2.5 and 3.0 T.}
\label{autonum}
\end{figure}

\begin{figure}
 \begin{center}
  \epsfxsize=14cm
  \epsfbox{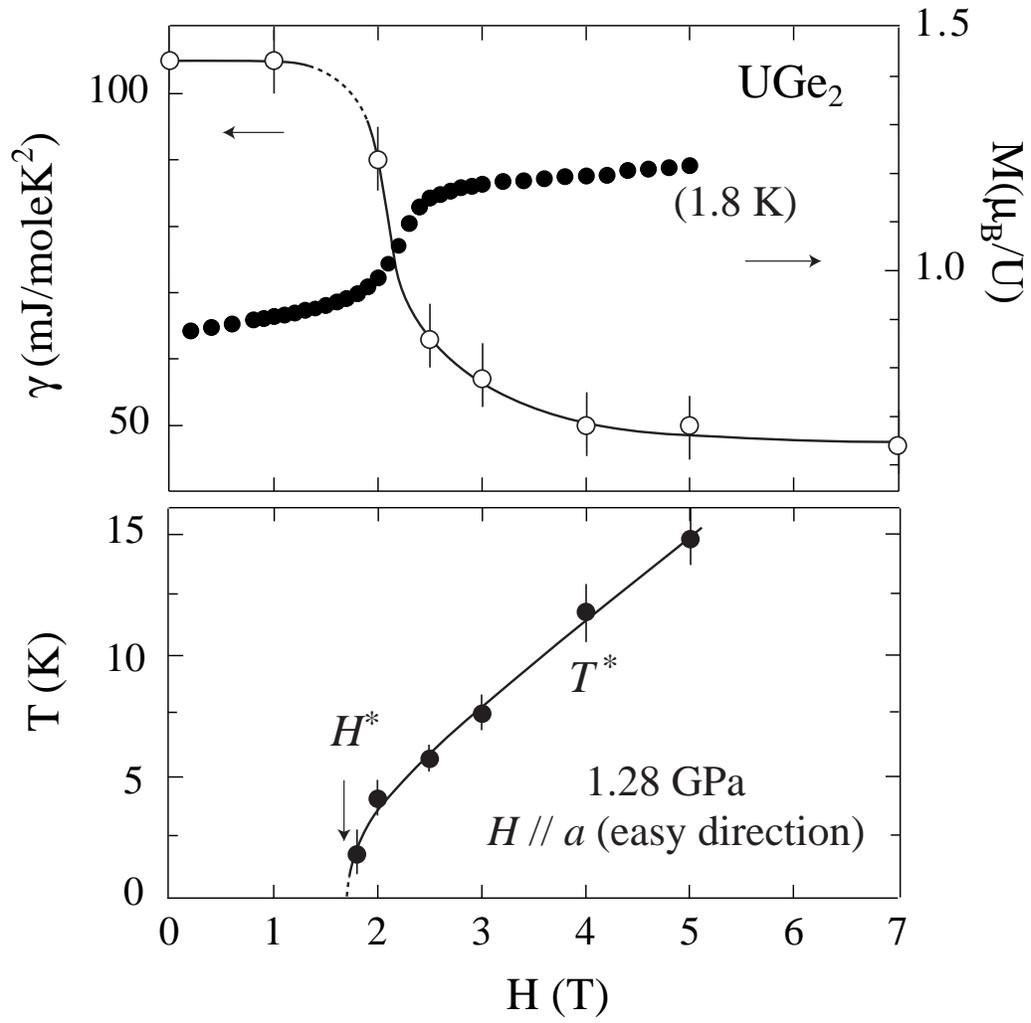}
 \end{center}
\caption{ Magnetic-field dependence of the magnetization, the linear heat-capacity coefficient $\gamma$, and $T^{*}$ for 1.28 GPa.}
\label{autonum}
\end{figure}
\begin{figure}
 \begin{center}
  \epsfxsize=14cm
  \epsfbox{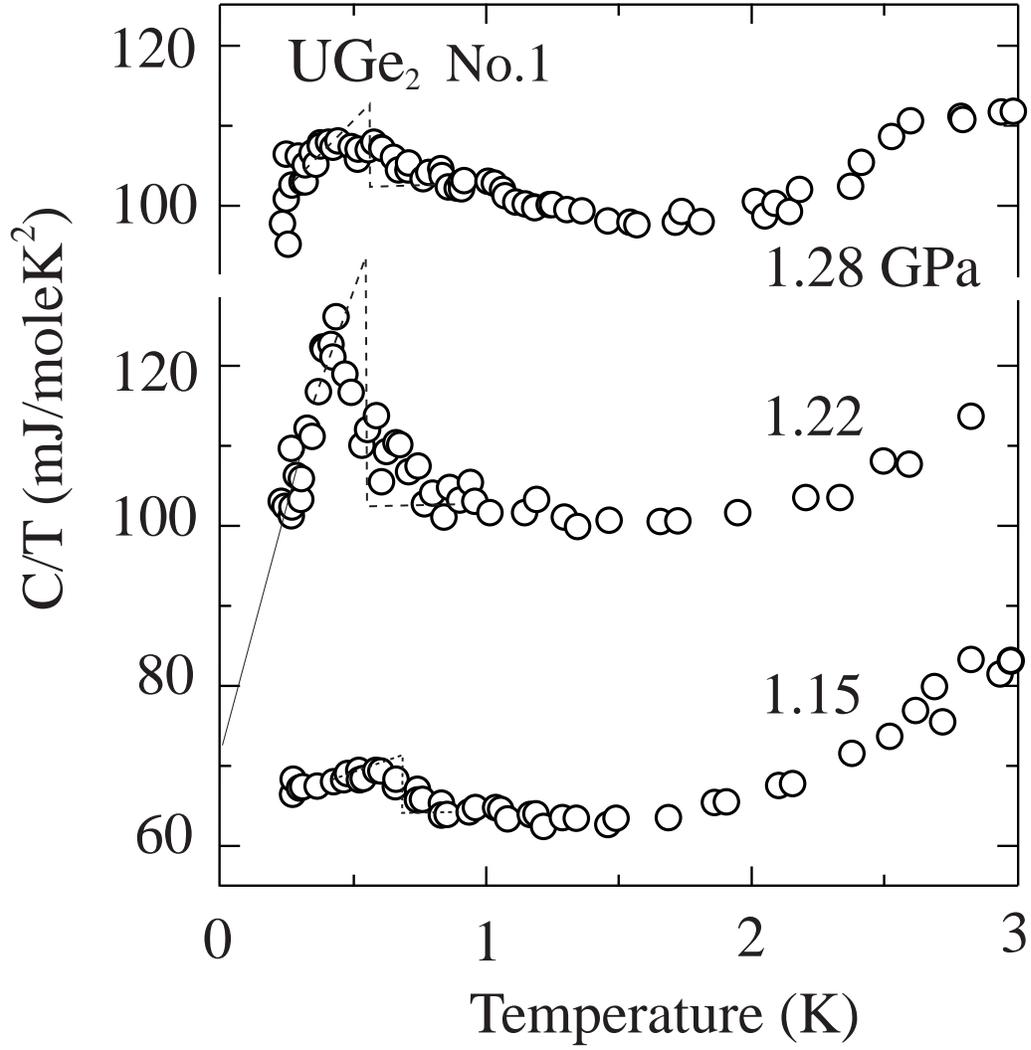}
 \end{center}
\caption{Temperature dependence of the heat capacity for the sample no. 1 at 1.15, 1.22, and 1.28 GPa. The dotted line represents an equal entropy construction at each pressure. The full line is the extrapolation of $C/T$ curve linearly to 0 K.}
\label{autonum}
\end{figure}

\begin{figure}
\begin{center}
  \epsfxsize=14cm
  \epsfbox{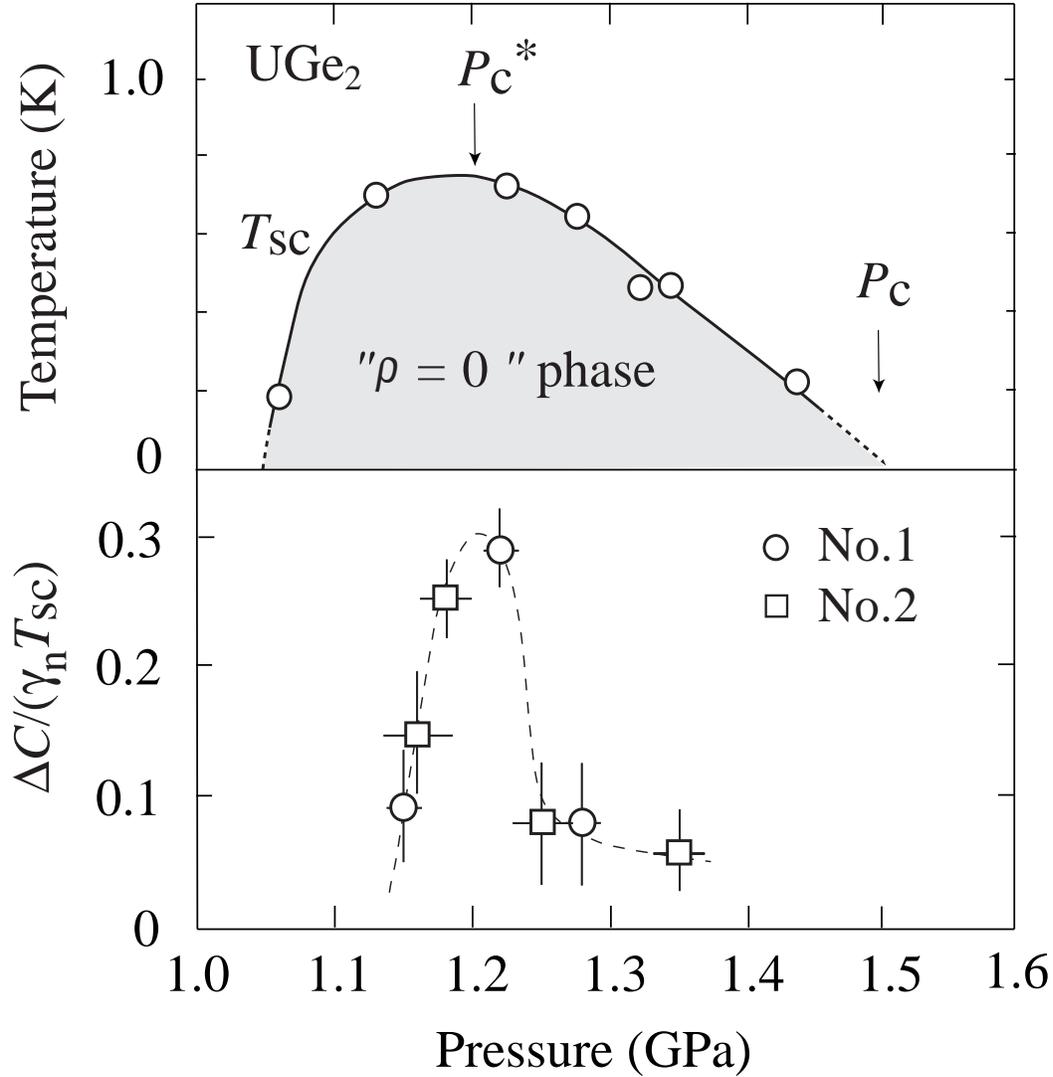}
\end{center}
\caption{ Pressure dependence of the superconducting transition temperature $T_{sc}$ determined by the resistivity measurement (the upper panel) and the value of ${\Delta} C/(\gamma_{n} T_{sc})$ (the lower panel). The full and dotted lines are guide for the eyes.}
\label{autonum}
\end{figure}

\end{document}